\documentclass[pss]{wiley2sp} 
\usepackage{amsfonts,amsmath,bm}
\newcommand{\bk}{b_{\bm{k}}}

\newcommand{\bkd}{b_{\bm{k}}^{\dag}}

\newcommand{\sumk}{\sum_{\bm{k}}}

\newcommand{\wk}{\omega_{\bm{k}}}

\newcommand{\kk}{\bm{k}}

\newcommand{\rl}{\rangle\!\langle}

\begin{document}

\title{Adiabatic rapid passage in quantum dots:
phonon-assisted decoherence and biexciton generation
}

\titlerunning{Adiabatic rapid passage in quantum dots:
phonon-assisted decoherence and biexciton generation
}

\author{%
	K. Gawarecki\textsuperscript{\textsf{\bfseries 1,2}},
	S. L\"uker \textsuperscript{\textsf{\bfseries 2}},
	D. E. Reiter\textsuperscript{\textsf{\bfseries 2}},
	T. Kuhn\textsuperscript{\textsf{\bfseries 2}},
	M. Gl\"assl\textsuperscript{\textsf{\bfseries 3}},
	V. M. Axt\textsuperscript{\textsf{\bfseries 3}},
	A. Grodecka-Grad\textsuperscript{\textsf{\bfseries 4}},
	P. Machnikowski\textsuperscript{\textsf{\bfseries 1\Ast}}
}

\authorrunning{K. Gawarecki et al.}

\mail{e-mail
  \textsf{Pawel.Machnikowski@pwr.wroc.pl}}

\institute{%
 \textsuperscript{1}\,Institute of Physics, Wroc{\l}aw University of
Technology, 50-370 Wroc{\l}aw, Poland\\
 \textsuperscript{2}\,Institut f\"ur Festk\"orpertheorie, Universit\"at M\"unster, Wilhelm-Klemm-Str. 10, 48149 M\"unster, Germany\\
  \textsuperscript{3}\,Theoretische Physik III, Universit\"at Bayreuth, 95440 Bayreuth\\
\textsuperscript{4}\,Niels Bohr Institute, University of Copenhagen,  2100 Copenhagen \O, Denmark
}

\received{XXXX, revised XXXX, accepted XXXX} 
\published{XXXX} 


\abstract{%
%
%
%
\abstcol{%
We study the evolution of a quantum dot controlled by a frequency-swept (chirped), linearly polarized laser pulse in the presence of carrier-phonon coupling. The final occupation of the exciton state is limited both due to phonon-induced transitions between the adiabatic spectral branches and because of phonon-assisted transitions to the biexciton state.  }
{
When the biexciton shift is large enough, the quantum dot can be modeled as a two-level system, which corresponds to excitation with circularly polarized light. For this case, we compare different methods of simulations: (i) a time convolutionless method, (ii) correlation expansion and (iii) path integrals.  We show that results obtained from these methods agree perfectly at low temperatures.}}

\maketitle   

\section{Introduction}
The interest in the optical control of the exciton states in a quantum dot (QD) has been stimulated by recent experiments, where the QD state was controlled either by means of Rabi oscillations \cite{ramsay10,ramsay10b} or by using chirped laser pulses exploiting the adiabatic rapid passage (ARP) \cite{simon11,wu11}. The usage of chirped pulses offers the possibility of a quantum control, which is not sensitive to small variations of the pulse area as Rabi oscillations are. In both cases, the role of phonons has been shown to be crucial, because their interaction with the QD leads to a reduction of the control efficiency \cite{forstner03,luker12,reiter12,roy11,debnath12,creatore12,gawarecki12b,glaessl13}. For the ARP, it was shown that at low temperatures the influence of phonons depends on the sign of the chirp. For negative chirps the control is limited by phonon interaction while for positive chirps no damping is seen for an excitation with circularly polarized light \cite{luker12}. When under linearly polarized excitation a phonon-assisted transition to the biexciton is possible thus hinders the control of the exciton state also for positive chirps \cite{gawarecki12b}.

To describe the phonon influence on the ARP, many theoretical approaches have been used like correlation expansion \cite{luker12,reiter12}, time-convolutionless approach \cite{gawarecki12b}, path integrals and the Bloch-Redfield-Wangsness theory \cite{eastham12}. Though the methods differ in their complexity, they all show a phonon-induced damping. In this paper, we want to make a quantitative comparison between three of those methods: (i) correlation expansion (CE), (ii) time-convolutionless method (TCL) and (iii) path integrals (PI). We furthermore discuss when the limiting case of a two-level model is a good assumption.

\section{Model}
We consider a QD modeled as a four-level system driven by a laser pulse with either circular or linear polarization. The Hamiltonian of the excitonic system is
\begin{displaymath}
H_{X}=E_{0}\left( |\sigma_{+}\rl \sigma_{+}| + |\sigma_{-}\rl \sigma_{-}| \right)
+(2E_{0}+E_{\mathrm{B}})|B\rl B|.
\end{displaymath}
$|0\rangle$ denotes the empty dot, $|\sigma_{\pm}\rangle$ are the two exciton states with different circular polarizations, $|B\rangle$ represents the biexciton state, $E_{0}$ is the single-exciton transition energy and $E_{\mathrm{B}}$ is the biexciton shift. The states with circular polarization  $|\sigma_{\pm}\rangle$ are related to the `linearly polarized` states $|X\rangle$,$|Y\rangle$ by $|\sigma_{\pm}\rangle=(| X \rangle \pm i |Y \rangle)/\sqrt{2}$. 
A Gaussian laser pulse with initial pulse area $\Theta$ and initial pulse duration $\tau_{0}$ is frequency modulated by a chirp filter with coefficient $\alpha$. The envelope
of the chirped pulse is then
\begin{equation*}
\Omega(t)=\frac{\Theta}{\sqrt{2\pi\tau_{0}\tau}}
\exp\left(-\frac{t^{2}}{2\tau^{2}}\right),
\label{Omega}
\end{equation*}
with the duration $\tau=(\alpha^{2}/\tau_{0}^{2}+\tau_{0}^{2})^{1/2}$ and the frequency chirp rate $a=\alpha/(\alpha^{2}+\tau_{0}^{4})$ \cite{luker12}. The central frequency of the laser pulse $\omega_{0}$ is tuned to the polaron shifted single exciton resonance.

In the rotating wave approximation, the Hamiltonian describing a carrier-light interaction for $\sigma_{+}$ circular polarization takes the form
\begin{displaymath}
H^{\mathrm{(circ)}}_{\mathrm{las}}=\frac{\hbar \Omega(t)}{2}
\left( |0 \rl \sigma_{+} | + |\sigma_{-} \rl B|  \right)
 e^{i\omega_{0}t+iat^{2}/2} +\mathrm{h.c.}
\end{displaymath}
Because the ground state is coupled only to $|\sigma_{+}\rangle$, the system can be reduced to a two-level model. For $X$ linear polarization the Hamiltonian reads
\begin{displaymath}
H^{\mathrm{(lin)}}_{\mathrm{las}}=\frac{\hbar \Omega(t)}{2}
\left( |0 \rl X | + |X \rl B|  \right)
 e^{i\omega_{0}t+iat^{2}/2} +\mathrm{h.c.}
\end{displaymath}
Here the coupling is $|0\rangle\leftrightarrow|X\rangle\leftrightarrow|B\rangle$. The other state ($|Y \rangle$) is uncoupled, therefore only three-levels are needed \cite{schmidgall10}. Finally, in order to simplify the numerical simulations, a unitary transformation to the non-uniformly rotating frame is made \cite{gawarecki12b}.

For a neutral exciton, large degree of charge cancellation reduces the piezoelectric coupling. Therefore, for the carrier-phonon coupling, we take into account the deformation potential interaction which couples the exciton to longitudinal
acoustic phonon reservoir. For the typical parameter range used in experiments \cite{simon11,wu11}, the generation of optical phonon modes is unlikely and can be neglected. The carrier-phonon Hamiltonian is
$H_{\mathrm{int}} =S\otimes R$,
where
\begin{displaymath}
 S = |\sigma_{+} \rl \sigma_{+} | + | \sigma_{-}  \rl \sigma_{-} | +2 |B \rl B|
\end{displaymath}
and
\begin{displaymath}
 R= \sum_{\kk} g_{\kk} \bk + \mathrm{h.c.}.
\end{displaymath}
$g_{\kk}$ are the coupling matrix elements \cite{gawarecki12b} and $\bk$ are the phonon annihilation operators. The energy of free phonons is described by $H_{\mathrm{ph}}=\sumk\hbar\wk\bkd\bk$, where $\wk=ck$ are the phonon frequencies. We consider a spherical QD with a size of $5$~nm. All of the parameters have been taken from \cite{luker12}.

We perform the simulations using three different methods. Because all of the methods are well described in the given references, we just state shortly the main aspect of the different approaches: (i) In the TCL method, which we implement in the 2nd order in the phonon coupling, only the reduced density matrix is represented in the simulation and, in addition, no explicit memory structure is present  \cite{breuer02,machnikowski08c,gawarecki12b}. (ii) The correlation expansion contains higher order density matrices, in particular, phonon related variables \cite{krugel05,krummheuer02}. We include terms up to 4th order in the phonon coupling. (iii) The PI method directly calculates the time evolution operator at discrete times \cite{vagov11a,vagov11b}. The obtained solution is numerically exact and thus provides a benchmark for approximate methods like the TCL or the CE. For the Rabi oscillations we already showed that the CE and PI agree perfectly for sufficiently low temperatures \cite{glaessl11}.

\section{Results}
\begin{figure}[tb]
\begin{center}
\includegraphics[width=85mm]{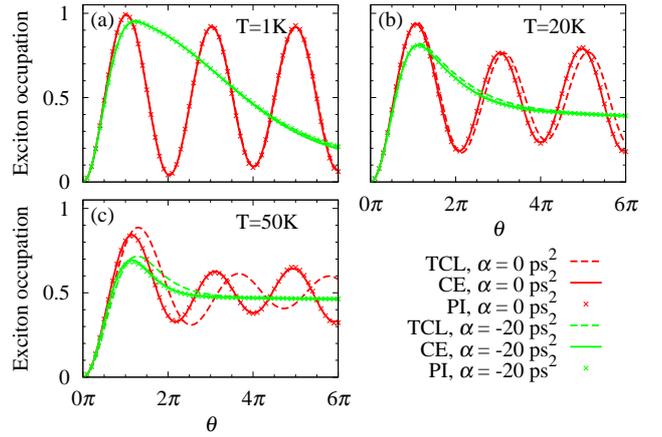}
\end{center}
\caption{\label{fig:cs}(Color online) Final occupation of the single-exciton state as a function of initial pulse area $\Theta$ at three different temperatures: (a) $T=1$~K, (b) $T=20$~K and (c) $T=50$~K.
Red lines show results for $\alpha = 0$~ps$^2$, green ones for $\alpha = -20$~ps$^2$.
Dashed lines are data obtained from the TCL, solid lines represent CE and points show results from PI.
 }
\end{figure}

\begin{figure}[tb]
\begin{center}
\includegraphics[width=75mm]{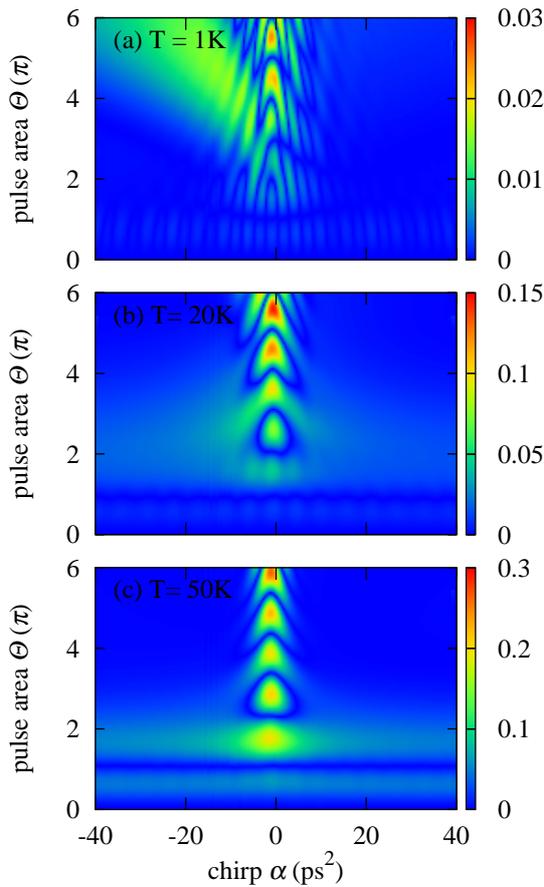}
\end{center}
\caption{\label{fig:map}(Color online) Absolute value of the difference between the final occupation of the single-exciton state calculated using TCL and CE method as a function of the initial pulse area $\Theta$ and the chirp $\alpha$ at: (a) $T=1$~K, (b) $T=20$~K and (c) $T=50$~K.  }
\end{figure}

\begin{figure}[tb]
\begin{center}
\includegraphics[width=85mm]{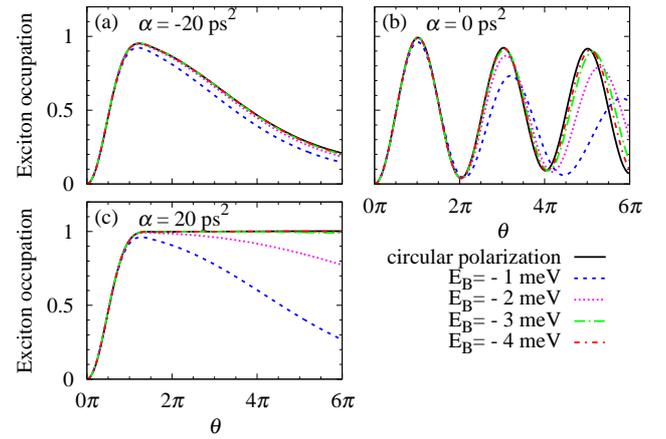}
\end{center}
\caption{\label{fig:biex}(Color online)
Final occupation of the single-exciton state as a function of the initial pulse area
$\Theta$ for different binding energies $E_B$ at $T=1$~K at three different chirp values: (a) $\alpha=-20$~ps$^{2}$, (b) $\alpha= 0$~ps$^{2}$ and (c) $\alpha=20$~ps$^{2}$.
}
\end{figure}

We start the discussion with the two-level model, where the QD is excited by circularly polarized light. The simulated evolution of the exciton occupation obtained by the three different methods for two different chirps $\alpha=-20$ and $0$~ps$^{2}$ is shown in Fig.~\ref{fig:cs}. For zero chirp we find the well known Rabi oscillations, which are damped by phonon interaction. For $\alpha=-20$~ps$^2$ the ARP leads to an occupation for the exciton state, which is also damped by phonon interaction. For increasing temperature the damping becomes stronger. 

For both values of $\alpha$ at $T=1$~K, the results show an excellent agreement of the different simulation methods (Fig.~\ref{fig:cs}(a)). At a moderate temperature of $T=20$~K the accordance between the TCL and the CE is also reasonable as shown in Fig.~\ref{fig:cs}(b). Note, that the mismatch between the TCL and the CE is larger for Rabi oscillations ($\alpha=0$~ps$^{2}$) than for the ARP ($\alpha=-20$~ps$^{2}$). At $T=50$~K the mismatch is getting higher (Fig.~\ref{fig:cs}(c)). For the Rabi oscillations a renormalization of the oscillation period due to the carrier-phonon interaction occurs. This is not well described in the TCL, such that the oscillation periods obtained by the two methods do not agree anymore leading to the observed difference. For the ARP at $\alpha=-20$~ps$^{2}$, these renormalizations are not so important and the results agree reasonably well even for this temperature. These approximate methods can be tested by comparing the results to the data obtained by PI. We find that results obtained from the CE and the PI agree perfectly for all temperatures considered in this paper.

In order to investigate quantitatively the mismatch between the TCL and the CE methods, we calculated the absolute value of their difference in a wide range of parameters. Fig. \ref{fig:map} presents these results as a map. As it can be seen in Fig. \ref{fig:map}(a), at $T=1$~K, the absolute value of the difference is below $3$~\%. Furthermore, in this case, the mismatch is higher for negative values of chirp. The reason is, that at low temperatures, for negative values of the chirp the correction to the final occupation from carrier-phonon interaction is much higher than for positive one \cite{gawarecki12b}. As shown in Fig.~\ref{fig:cs}(b), the discrepancy at $T=20$~K can be up to 15\% for the small chirp $\alpha$. Fig.~\ref{fig:cs}(c) shows, that at high temperatures ($T=50$~K), near the Rabi oscillation region, the mismatch between methods can reach 30\%. As already seen in Fig.~\ref{fig:cs}(c), the large difference between TCL and CE is mostly related to the difference in oscillation periods for the Rabi oscillations. For chirped excitation this renormalization does not play a big role, such that here the difference between the two methods is rather small.

In the next step, we consider a linearly polarized laser pulse. We investigate the influence of the biexciton shift $E_{\mathrm{B}}$ on the decoherence of the single-exciton state. We increase the value of $E_{\mathrm{B}}$ successively from $E_{\mathrm{B}}=-1$~meV to $E_{\mathrm{B}}=-4$~meV and compare these results with the ones obtained from the two-level system (for circularly polarized light), which should agree for very high biexciton shifts $E_B$ \cite{gawarecki12b}. The final occupation as a function of the pulse area for three different chirp values is presented in Fig.~\ref{fig:biex}. For $\alpha=-20$~ps$^{2}$ in Fig.~\ref{fig:biex}(a) the phonon coupling to the exciton state dominates the damping of the ARP and phonon-induced transitions to the biexciton state play a minor role even for the very small biexciton shift $E_{\mathrm{B}}=-1$~meV. In consequence, the discrepancy of the single-exciton occupation between the two- and four-level system is small. Fig.~\ref{fig:biex}(b) shows that in the case of Rabi oscillations at $\alpha=0$~ps$^{2}$ for small biexciton shifts an additional damping related to phonon-assisted biexciton transitions appears. In particular, for high pulse areas a strong deviation from the two-level model is observed. For $\alpha=20$~ps$^{2}$ shown in Fig.~\ref{fig:biex}(c) no dephasing due to phonons takes place in the case of the two-level system. Here, for the small biexciton shifts $E_{\mathrm{B}}=-1$~meV and $E_{\mathrm{B}}=-2$~meV, the phonon-assisted biexciton generation leads to a decrease of the exciton occupation for high pulse areas.

For all chirp values, we find that for a biexciton shift of $E_{\mathrm{B}}=-4$~meV the four-level model agrees with the two-level model. This is because a biexciton shift of $E_{\mathrm{B}}=-4$~meV leads to an energy separation above the cut-off frequency of the carrier-phonon coupling and thus no phonon-assisted biexciton generation is possible. The same effect takes place for the positive biexciton shift. The results for $E_{\mathrm{B}}=2$~meV (not shown in the pictures) show perfectly agreement with the case of circular polarization.

\section{Summary}

In summary, we have investigated the exciton dynamics in a QD adiabatically controlled by a chirped laser pulse. For circular polarization we have compared results obtained from three different simulation methods namely (i) TCL, (ii) CE and (iii) PI. The results obtained from CE and PI show very good agreement, even at $T=50$~K. The TCL agrees perfectly with the other two methods at low temperatures and agrees reasonably at moderate temperatures, but at high temperatures a mismatch of up to 30\% was found. The agreement between these methods is better for the ARP dynamics than for the Rabi oscillations. We have also studied exciton dynamics for a linearly polarized laser pulse. We have shown, that new decoherence paths related to phonon-assisted biexciton generation open up, which are efficient for biexciton shifts of the order of $-2$~meV. For higher biexciton shifts the influence of the biexciton becomes negligible and the system can be reduced to a two-level model. 

This work was supported in part by a Research Group
Linkage Project of the Alexander von Humboldt Foundation
and by the TEAM programme of the Foundation for Polish
Science, cofinanced from the European Regional Development
Fund.

%

\providecommand{\WileyBibTextsc}{}
\let\textsc\WileyBibTextsc
\providecommand{\othercit}{}
\providecommand{\jr}[1]{#1}
\providecommand{\etal}{~et~al.}

\end{document}